# Review of algorithms for predicting fatigue using EEG

Ildar Rakhmatulin, PhD

Miruns, miruns.com ,

email: ildarr2016@gmail.com

**Abstract**
Fatigue detection is of paramount importance in enhancing safety, productivity, and well-being across diverse domains, including transportation, healthcare, and industry. This scientific paper presents a comprehensive investigation into the application of machine learning algorithms for the detection of physiological fatigue using Electroencephalogram (EEG) signals. The primary objective of this study was to assess the efficacy of various algorithms in predicting an individual's level of fatigue based on EEG data.

**1. Introduction**
Since 1878 the French physiologist Angelo Mosso [13] has carried out pioneering studies of the blood circulation in the brain during mental and physical work, initiating an understanding of the physiological basis of fatigue and the study of physiological fatigue, research efforts have already spanned several disciplines, including psychology, physiology, neurology and occupational health. Over the years, scientists and researchers have made significant contributions to understanding the nature, causes, and consequences of physiological fatigue. The prediction of physiological fatigue is critical in areas where performance, safety, human well-being and especially sports are of paramount importance. By understanding and predicting fatigue levels it is possibly take proactive steps to reduce fatigue-related risks, optimize performance, and improve overall health and safety.
Today, the prediction of fatigue is required in huge quantities of areas from the prediction of the tiredness of the driver, which is one of the main factors related to traffic accidents [14] and ending with sports implementations. For example, it is known that several players of the Italian football team, who won the 2006 World Cup, used a neurobio management to improve their results [15]. More details about the history and physiology of fatigue, and the feasibility of its detection are described in the work [16, 17]. There are also a large number of survey works in which the authors examined various factors affecting fatigue [18], [19]. Hooda et al. [20] presented this topic more locally and considered the machine learning techniques for detecting fatigue and Wang et al. [21] did similar work but with an emphasis on the EEG signal.
In recent years, Brain-Computer Interfaces (BCIs) have become widely available in the market [36, 37, 40, 41, 42]. Simultaneously, the use of machine learning algorithms for feature extraction from EEG signals has gained popularity across various neuroscience domains [38, 39, 43, 44]. One such promising application area is the detection of fatigue through EEG signals.
Currently, there are comprehensive works [32] that offer a review of different approaches for detecting fatigue using machine learning techniques. Additionally, various studies [33] provide an overview of devices capable of determining fatigue. Notably, articles [34, 35] present detailed reviews of available low-cost EEG headsets used for drowsiness detection, which are closely related to this research.
In light of the existing literature, this work represents a significant advancement as it specifically focuses on machine-learning algorithms for fatigue detection.

**2. Short introduction to EEG**
Measure EEG it is a non-invasive method used to measure the electrical activity of the brain. Brain cells communicate with each other using electrical impulses, and the EEG records these electrical signals using several electrodes placed on the scalp. During an EEG, the electrodes detect electrical activity generated by neurons in the brain. These electrical signals are amplified and recorded, creating a graphical representation called an electroencephalogram. An EEG recording shows patterns of electrical activity known as brain waves, which can provide valuable information about brain function and activity. For example, delta waves occur in the 0.5 Hz to 4 Hz frequency range and are present during deep sleep, while beta waves occur in the 13 Hz to 30 Hz range and are associated with active thinking. Similarly, other waves are associated with - alpha waves (8-12 Hz): normal waking conditions, gamma waves (30-80 Hz): integration of sensory perception and theta waves (4-7 Hz) [22].

### 3. What is fatigue?

Fatigue can manifest itself in different ways, in this paper we focus on physiological fatigue, but at the same time, we will mentions other types of fatigue since they are correlated with each other:

- **Physical fatigue.** Physical fatigue is the best known type and is often caused by prolonged physical activity or exertion. This can be caused by factors such as muscle fatigue, lack of physical fitness, or overexertion;
- **Mental fatigue.** Mental fatigue is associated with cognitive activity and prolonged mental effort. This may be the result of tasks that require concentration, problem solving, or intense concentration. Mental fatigue can affect cognition, concentration, and decision making. Leshko [8] et al detected Influences of mental fatigue on physical fatigue;
- **Emotional fatigue.** Emotional fatigue is characterized by a feeling of emotional exhaustion and is often associated with prolonged periods of stress, anxiety, or emotional tension. This can be caused by factors such as relationship problems, stress at work, or personal issues. **Emotional fatigue** also has Impact on fatigue [23];
- **Chronic Fatigue:** Chronic fatigue refers to persistent and prolonged fatigue that is not relieved by rest or sleep. This is often accompanied by other symptoms such as weakness, memory problems, and difficulty concentrating. Chronic fatigue syndrome (CFS) is a condition that causes extreme and unexplained fatigue and obviously has impact to fatigue [24].

But at the same time, fatigue is affected by a huge number of factors, for example, stress affects fatigue [25] and there are a number of works that demonstrate how to detect stress through EEG [26]. Fatigue is affected by drowsiness [27] which can be easily detected through the EEG [28] as well as insomnia [29]. The patterns can be very different, even in unexpected forms, for example, avoiding caffeine increases the rate of cerebral blood flow and changes the activity of quantitative electroencephalography (EEG), and the rate of cerebral blood flow affects the Middle cerebral artery blood velocity during running [30]. There is a huge variety of subtypes of not explicitly expressed patterns for Heavy legs, Breathing difficulty, Decreased motivation, Muscle cramps, and etc.

Most often, fatigue detection through EEG signals relies on analyzing the alpha rhythm. The alpha rhythm has consistently demonstrated a strong correlation between its signal strength and the subject's level of fatigue [11, 6]. Researchers commonly utilize the signal power spectral density to investigate this correlation. For instance, in the paper [2], the authors applied a non-linear method to analyze the power spectral density of EEG signals. To process the EEG data, they first applied a 0.5 Hz high-pass filter and a 30 Hz low-pass filter to attenuate the signals. Next, independent component analysis (ICA) was employed to remove artifacts such as eye movements, myoelectricity, noise, and other disturbances. On Fig. 1 showed the constructed EEG spectrum and a topographic map of the brain of three states.

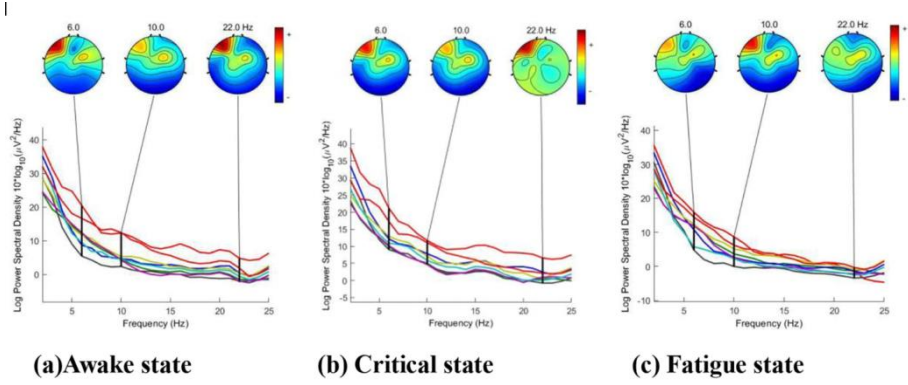

Fig. 1. EEG spectrum and brain topographic map of three states.

As can be seen from the spectrogram, during the transition from the state of wakefulness to the state of fatigue, the amplitude tends to smoothness. The amplitudes of the α- and β-waves decrease, while the amplitude of the θ-wave has no obvious changes.

## 4. Algorithms for determining fatigue

This research presents an extensive investigation of various simple techniques, namely linear, multi-level complex work, Q learning, and spindle paradigm. Our objective is to elucidate the individual merits and applications of these approaches. By dividing the study into distinct points, we provide a structured and systematic analysis of each technique, enabling a deeper understanding of their respective capabilities and limitations.

### 4.1 Linear Algorithms to predict fatigue

The implications of the findings are discussed in this section, including the practical applicability of the identified linear algorithm in real-world occupational settings. In the paper [4], the authors used a regression method for EEG-based fatigue detection using cross-sectional data, 23 subjects, and 12-channel EEG signals. In this study, the authors filtered the data set with 1 Hz high-pass and 50 Hz low-pass filters with finite impulse response (FIR), Fig.2.

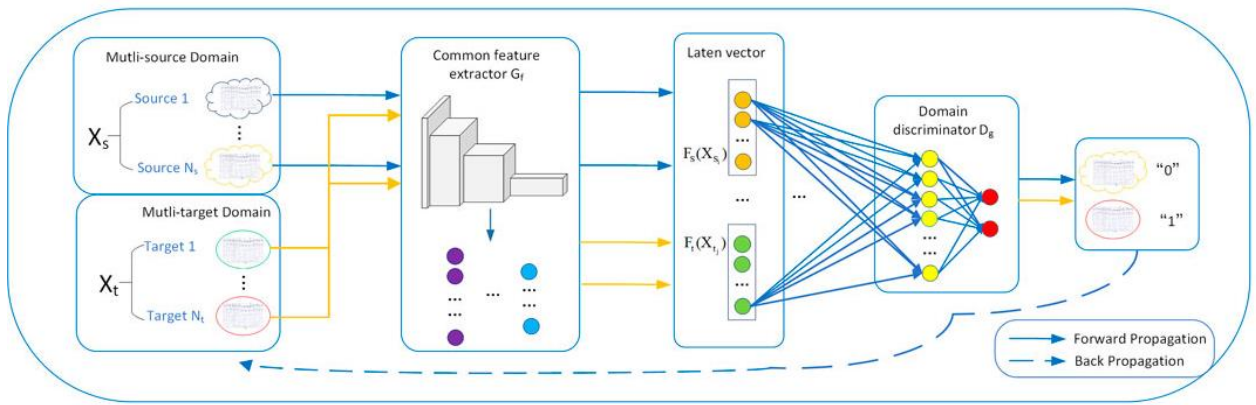

Fig.2. Regression Method for EEG-Based Fatigue Detection

This study showed that the performance of shallower models saturates faster to lower performance levels compared to deeper networks. In this study, ResNet50 was chosen as the CNN to extract the spatial local features of univariate EEG samples. The entire subject response time (RT) to an event in the experiment was collected. 5% of the shortest and fastest RT of each subject were extracted, and the sorted RT difference for the shortest and fastest RT was calculated as the behavioral outcome. The EEG data for the shortest and fastest RT were collated to discuss different brain dynamics in the two behavioral settings.

After the behavioral outcome, the frequency power of the subject was calculated to know the dynamic changes in his brain. In this study, frequency power was divided into four ranges: delta (1 Hz), theta (4–8 Hz), alpha (9–12 Hz), and beta (13–30 Hz). The authors plotted the power of each band and sorted the RT, frequency power from 0 to 0 Hz of the subjects in the experiment, Fig.3.

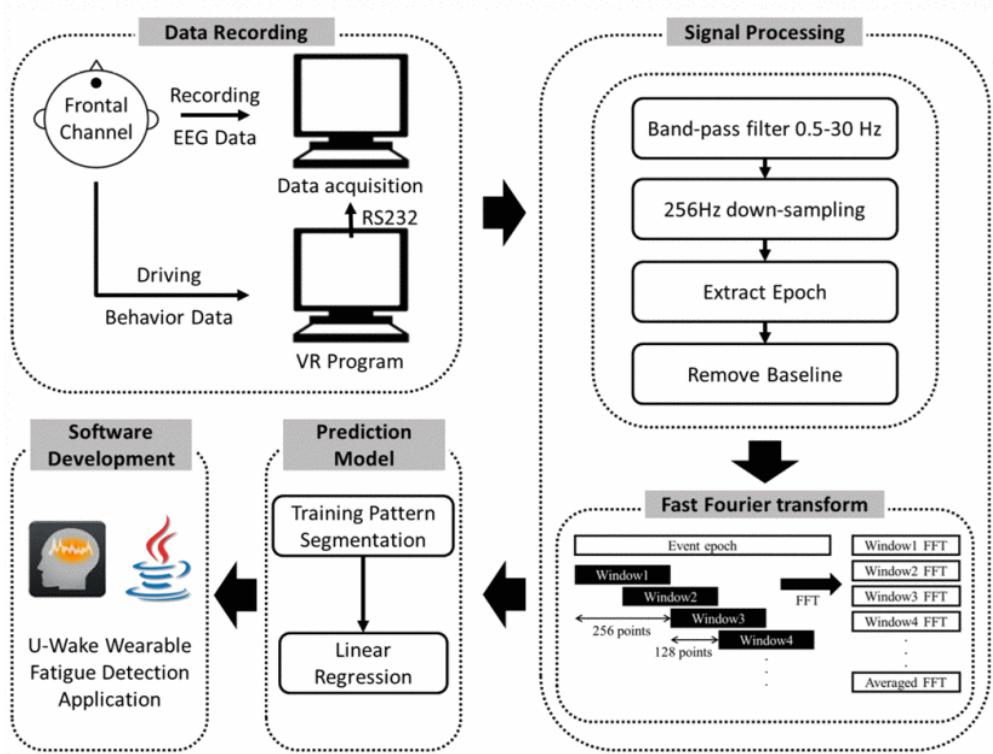

Fig.3. Block diagram of EEG signal processing and prediction regression models

The alpha range increase phenomenon was more evident than for other ranges, Fig.4.

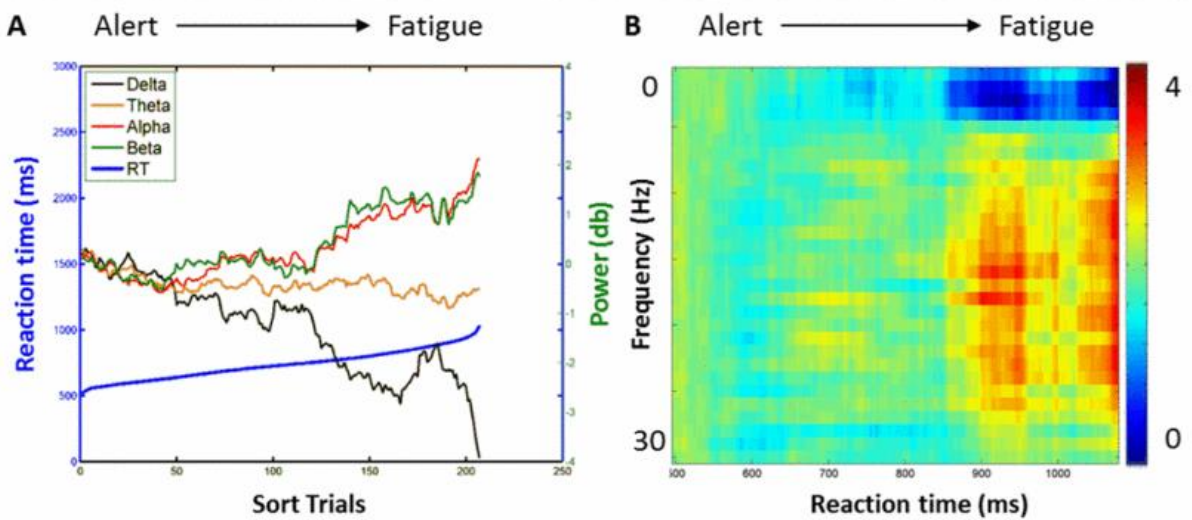

Fig. 4. (A) Delta, theta, alpha, and beta power spectra sorted by sorted RT. (b) Power spectrum image for one front channel.

The alpha range increase phenomenon was more evident than for other ranges. The prediction model was created through cross-validation with the exclusion of one subject. The mean accuracy in this study was 93.9%. With the exception of a few subjects, the accuracy of most of the subjects exceeded 90%, and the maximum accuracy could reach 99.7%. This showed that the regression model in this study was a promising method for predicting user fatigue levels.

**4.2 CNN to predict fatigue**

This section introduces how CNNs can be adapted for sequential data processing for fatigue prediction using physiological time-series data.

In the paper [5], a spectrogram of EEG signals was used as input to the CNN architecture. Mel's spectrogram is injected into the fatigue detection model to perform the fatigue detection task for EEG signals. The fatigue detection model consists of a 6-layer Convolutional Neural Network (CNN), Bidirectional Recurrent Neural Networks (Bi-RNN), and a classifier. In the modeling phase, spectrogram features are passed to a 6-layer CNN to automatically learn high-level features, thereby extracting temporal features in a bi-directional RNN to obtain temporal information about the spectrogram. The state of readiness or fatigue is obtained using a classifier consisting of a fully connected layer, the ReLU activation function and the softmax function, Fig.5.

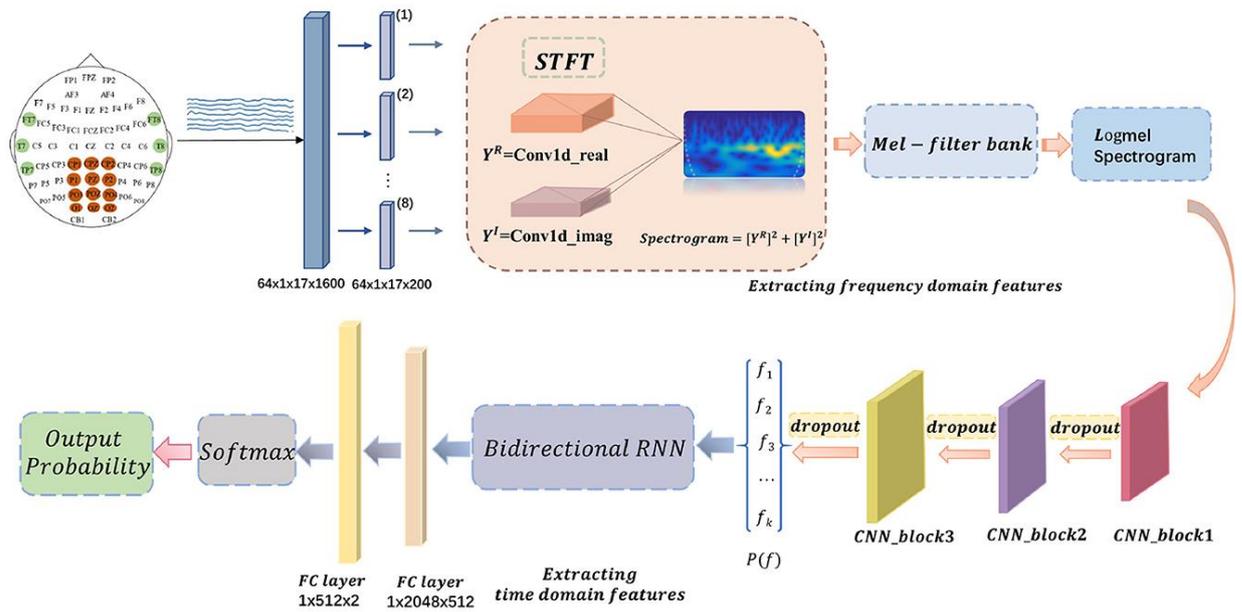

Fig.5. Schematic diagram of the entire model of the proposed LogMel-CRNN model

Another promising direction is to use Recurrent Neural Networks (RNNs) to predict fatigue via EEG. RNNs are a type of neural network architecture that can be well-suited for processing time-series data like EEG signals. There are several reasons why RNNs are considered advantageous for predicting fatigue via EEG: temporal dependencies, Variable Length Sequences, Long Short-Term Memory (LSTM) Cells

In the paper[10], the authors provided recurrent self-evolving fuzzy neural work (RSEFNN) to increase the memory capacity for adaptive noise reduction in assessing the mental state of drivers while driving a car. Experimental results without using the ICA procedure show that the proposed RSEFNN model retains superior performance compared to current models.

**4.3. Cascade architecture**

In this chapter, we will consider cascade architecture, which consists of multiple stages, each designed to address specific aspects of fatigue prediction. Each stage processes the input data sequentially, refining the predictions at each step and passing relevant information to the subsequent stages.

In the paper [3], the MATCN-GT model is presented, which consists of a multiscale attention block of a temporal convolutional neural network (MATCN block) and a graph convolutional transformer block (GT block). Among them, the MATCN block extracts features directly from the original EEG signal without a priori information, and the GT block processes features of EEG signals between different electrodes. A multi-level attention module was also used to ensure that valuable information about electrode correlations was not lost. The Transformer module was added to the graph convolutional network, which captures the dependencies between remote electrodes.

At the same time, complex architects in some cases can show performance no higher than standard machine learning algorithms. For example, in the paper [12], authors used Particle Swarm Optimization-based Hidden Layer Optimization Extreme Learning Machine (PSO-H-ELM) for the prediction of fatigue. But ultimately, experiments showed that the PSO-H-ELM algorithm has an advantage of only about 4% compared to the average accuracy of the KNN algorithm and the SVM algorithm.

### 4.4 Reinforcement learning

Reinforcement Learning (RL) is a type of machine learning that involves training an agent to make decisions in an environment to maximize a cumulative reward. While RL can be applied to various domains, including robotics and game playing, it is not commonly used for predicting fatigue directly via Electroencephalography (EEG). In the paper [7], the Deep Q-Network in combination with RNN was used to predict fatigue, Fig. 6.

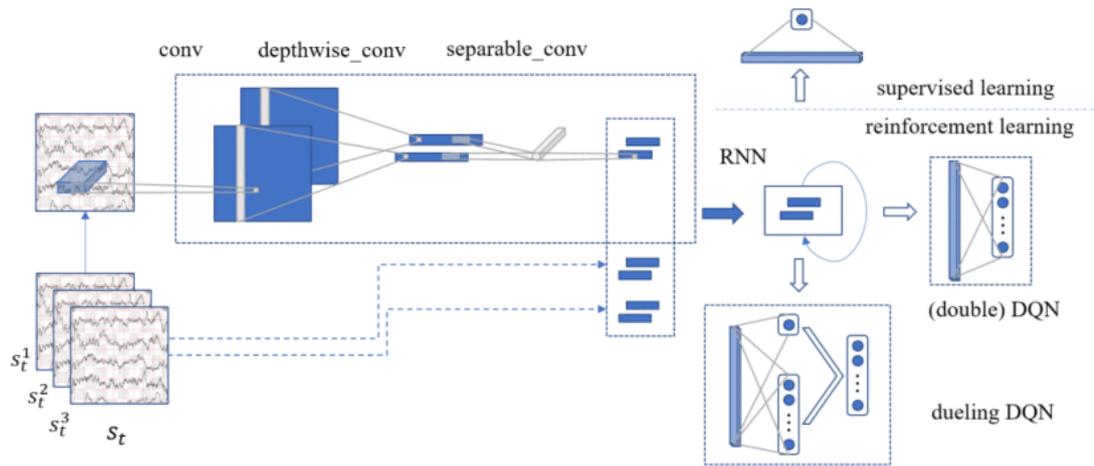

Fig.6. Architecture of Deep Q-Network and RNN

This architecture up to RNN can be used in both supervised learning and reinforcement learning paradigms. For reinforcement learning, DQN or double DQN can share the pre-RNN structure with the duel DQN.

### 5. EEG paradigms to predict fatigue

The EEG paradigm refers to a specific experimental design or task used in EEG research to investigate certain aspects of brain activity or cognitive processes. EEG paradigms are carefully structured to elicit specific brain responses, and these can vary widely depending on the purpose of the study.

In the paper [9], the authors presented an algorithm that identifies sudden bursts of narrow-band oscillatory activity in the EEG (spindles) using methods derived from the analysis of change points. Their motivating example is the discovery of alpha spindles in the parietal/occipital areas of the brain. Autoregression successfully identifies spindles that have a direct correlation with fatigue, Fig.7.

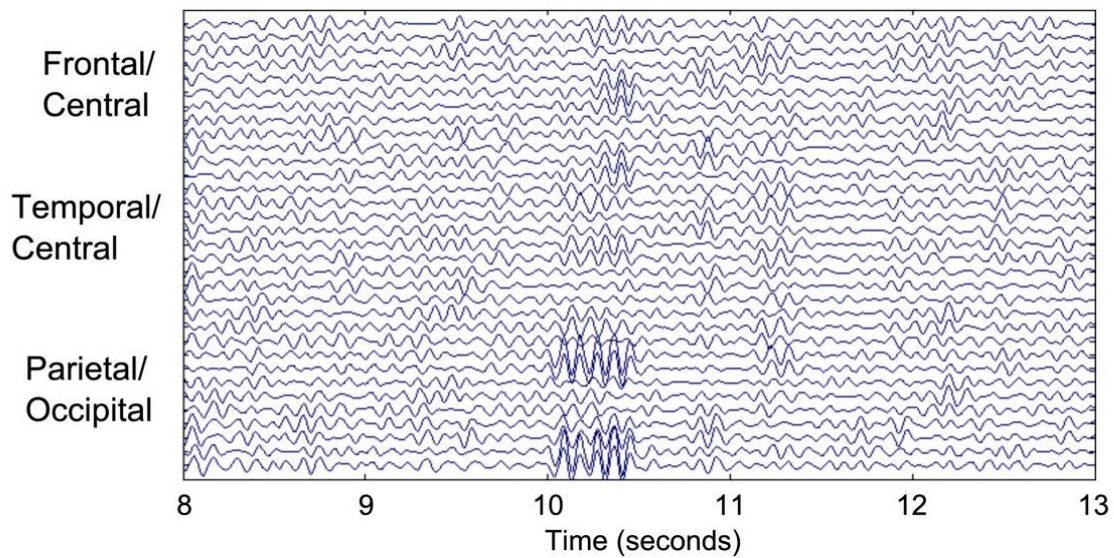

Fig.7. The simulated EEG activity, with the Y-axis denoting the channel locations, ordered left to right hemisphere, frontal to occipital.

**Conclusion**

Fatigue prediction plays a crucial role in detecting drowsiness in drivers and has gained significant traction in modern times. Numerous methods and approaches are employed to implement this idea, with key factors such as the number of EEG channels, electrode placement on the head, data reading frequency, signal quality, and the size of the subject dataset being essential indicators. However, dealing with EEG signals presents challenges due to their non-linear and non-stationary nature, influenced by various factors.

In existing studies, the preparation of EEG data typically involves band-pass filtering, downsampling, and artifact removal. Many research works have shown a correlation between signal strength in the alpha frequency band of EEG signals and the fatigue of the subject, often leading to binary classification of fatigue status. The input data for fatigue prediction tasks usually comprises spectral characteristics represented in 3D images for Convolutional Neural Networks (CNNs) and numerical values in time series for standard machine learning algorithms.

For more effective fatigue prediction, it is advisable to leverage recurrent neural networks (RNNs) that can capture temporal dependencies and possess memory. Researchers have successfully demonstrated the feasibility of using various cascade algorithms, ultra-precise networks, and standard algorithms like SVM to identify fatigue through EEG signals. Interestingly, the number of EEG channels has shown less influence on the accuracy of fatigue identification compared to its impact on other related tasks, such as stress level control or motor imagery.

However, implementing these algorithms in real applications faces challenges due to the extremely non-stationary nature of EEG signals. Future research efforts should focus not only on proving the ability of the presented algorithms to determine fatigue but also on establishing their universality and reliability under varying EEG signal measurement conditions and with diverse subjects. A critical requirement for achieving this is the availability of a large and diverse dataset. Nevertheless, collecting such data necessitates a wearable device capable of providing a comprehensive set of information about both the subject and the environment.

Unfortunately, wearable devices currently available in the market are application-specific and do not provide the required extensive data about the subject. However, the field of brain-machine interface development has been gaining popularity in recent years, which in stills optimism for the future of fatigue prediction research. Advancements in this area may lead to the creation of more versatile and comprehensive wearable devices, enabling the use of transfer learning mechanisms for the developed prediction models.